\documentclass[aps,notitlepage]{revtex4-1}
\usepackage{graphicx}
\usepackage{epstopdf}
\usepackage{verbatim}
\usepackage{amsmath}
\usepackage{amssymb}
\usepackage{graphicx}
\usepackage{subcaption}

\begin{document}

\title{The Kelly growth optimal strategy with a stop-loss rule}

\author{M.\ Nielsen\thanks{Email: mads.nielsen@glgpartners.com}}
\affiliation{GLG Partners LP, 1 Curzon Street, London W1J 5HB, UK}
\affiliation{Viking Science Ltd, London W8 7AY, UK.}

\begin{abstract}
From the Hamilton-Jacobi-Bellman equation for the value function we derive a non-linear partial differential equation for the optimal portfolio strategy (the dynamic control). The equation is general in the sense that it does not depend on the terminal utility and provides additional analytical insight for some optimal investment problems with known solutions. Furthermore, when boundary conditions for the optimal strategy can be established independently, it is considerably simpler than the HJB to solve numerically. Using this method we calculate the Kelly growth optimal strategy subject to a periodically reset stop-loss rule.
\end{abstract}

\maketitle
\tableofcontents

\section{Introduction}

The Kelly strategy is attractive to practitioners because of its robust and optimal properties, e.g.\ that it dominates any other strategy in the long run and minimizes the time to reach a target, as shown by \citet{Kelly56}, and among others by \citet{Breiman61}, \citet{Thorp69,Thorp71,Thorp06}, \citet{Hakansson70,Hakansson71}, \citet{Algoet88}, \citet{Merton90}, \citet{Browne00a,Browne00b}. It also carries intuitive appeal because of its connection to information theory, which was part of the original motivation in \citet{Kelly56}, and in particular by the fact that optimal investing is one and the same thing as optimal application of the available information; see \citet{Barron88} and \citet{Cover06}. On the other hand, the Kelly strategy in its pure unconstrained form, which we will refer to as the `free Kelly' strategy, generally leads to a too aggressive strategy with too high leverage and risk to be of practical applicability. Real investors are subject to various constraints, and a practical version of the Kelly principle can be formulated as: Optimize the objectively expected compound growth rate of the portfolio subject to the relevant constraints. It is in this form that the Kelly principle has become increasingly popular among practitioners over the last 20 years or so. In this paper we are concerned with the optimal strategy in this sense, for the particular case where the constraint is in the form of a stop-level. Unlike for most other forms of constraints, the effect of a stop-level on the optimal strategy varies with the portfolio value and with the time remaining until the stop-level is adjusted (reset). The related problem of finding the growth optimal strategy for a portfolio subject to a drawdown constraint (a trailing stop-loss rule) was solved using a combination of HJB and Martingale methods by \citet{Grossman93}, see also \citet{Cvitanic95,Cvitanic97} and section~\ref{GrossmanZhou} below.

The plan of the paper is as follows. Section~\ref{sec:moti} provides some background, and section~\ref{sec:hjb} reviews the HJB for the investment problem and introduces notation. In section~\ref{sec:nonlinPDE} we derive a non-linear PDE for the optimal strategy, and in section~\ref{sec:known} we consider a few examples with known solutions, and discuss to what extent the equation can be applied. In section~\ref{koswslr} we provide a numerical solution for the Kelly strategy subject to a periodically reset stop-loss rule. Section~\ref{sec:conclusion} concludes.

\section{Background and motivation}
\label{sec:moti}
Consider a discretionary portfolio manager running a trading book in a global macro hedge fund. Such a trader is subject to various constraints when selecting the appropriate level of risk to run. Apart from leverage, stress-test and concentration limits and maybe constraints on the asset selection, the book will typically be subject to a value-at-risk (VaR) limit. In practice most traders will run their book significantly below these limits because there is another kind of constraint: The stop level. The stop may take several forms. It is for example common for a book with monthly liquidity to be subject to explicit soft and hard stop levels between liquidity dates. A book with a $3\%$ VaR limit at $95\%$ 1~day, could for example have a soft stop at $-5\%$ MTD (month-to-date) where the VaR limit drops to $1\%$ and a hard stop at $-10\%$ MTD where the VaR limit drops to zero for the remainder of the month. In any given month the stop-level is a fixed number of percentage points below the portfolio value at the beginning of the month. Any triggered stop will be released and reset after the first coming month-end, where investors will be given a chance to withdraw their funds. Hired (non-partner) traders managing a relatively small sub-book as part of a larger hedge fund are typically also subject to an explicit terminal stop at around $-25\%$ YTD (year-to-date), at which point the book is closed and they lose their job. Even founders and senior partners, who may not be subject to any explicit stop levels, will in effect operate under an implicit stop level. This is due to the fact that if the drawdown from the high-water mark becomes too large, then there will be no performance fees for the foreseeable future to pay for the operation, and the fund is likely to find itself in a downward spiral of brain-drain, investor withdrawals, and lower returns. The desire to avoid this situation will in effect mean that the fund can only operate at full risk when it is near its high-water mark. Indeed, it has happened more than once that a hedge fund has simply closed after a steep drawdown, only for the managers to open a new fund in a different setting. History has of course also shown what can happen when a trading operation is run without stops or with unenforceable stops due to illiquidity or management failure. Leaving the moral aspects aside, it is clear that stop-levels, implicit or explicit, are a major part of the constraints limiting the risk in most funds, and that the constraining effect of the stop level scales with the portfolio value and the time remaining until the stop level is reset.

The penalty for hitting the stop is lost opportunity. For a terminal stop the opportunity cost is potentially infinite since the loss is all future earnings. A strategy can try to avoid hitting the stop by reducing the risk as the portfolio value approaches the stop level, but in practice it is impossible to guarantee that the stop will not be hit unless the risk is zero. In a model world where the risky assets are modelled using Brownian motion, it is possible to avoid hitting the stop by reducing the risk, but this does not mean that there is no opportunity loss. Scaling down the risk also scales down potential returns, and increases the amount of time the portfolio end up spending in the vicinity of the stop. Thus there is a kind of `dead zone' just above the stop level, where the portfolio has very little risk, and it can become virtually stuck. The optimal strategy must find the right balance between scaling the risk down too quickly versus too slowly, foregoing potential returns in both cases. With a terminal stop, or an infinitely long time to reset, the solution is to effectively ignore the capital below the stop level, and invest in the Kelly optimal way as if the capital above the stop level was the total capital. Doing this, the trader is only risking the fraction of the capital he can survive to lose. This strategy which is independent of time, but a function of the portfolio value is discussed in section~\ref{sec:kwts}, and the related optimal strategy for a portfolio subject to a trailing maximum drawdown limit is discussed in~\ref{GrossmanZhou}.

A stop level which resets at the end of each period is more interesting from a practical point of view. With a periodically reset stop-level the penalty for hitting the stop, or becoming stuck in the dead zone, is less severe, since it is only a temporary loss of opportunity. If there is only a few days left before the reset, the portfolio value must be very near the stop-level for this to have any effect, but in the other limit where there is a very long time to the reset, the optimal strategy must approach the same strategy as the one for a terminal stop/infinite horizon. It is therefore clear that an optimal strategy will scale the risk up and down, not only according to how far away the stop is from the current portfolio value, but also according to the time remaining until the stop-level reset. It is the Kelly growth optimal strategy for this problem, discussed in detail in section~\ref{koswslr}, which is our main objective.

A realistic modelling of the problem would be very complicated, depend on idiosyncratic details, and is beyond our ambitions in this paper. Instead our goal is to study the optimal strategy in the simplest non-trivial model with a stop-rule. As it turns out, the resulting strategy, discussed in section~\ref{koswslr}, corresponds quite closely to common trader intuition, and can provide a rough guide to the appropriate risk level for practical portfolios.

\section{The value function and the HJB equation}
\label{sec:hjb}
We use the standard minimal portfolio model consisting of a risky asset displaying geometric Brownian motion and an exponentially growing risk-free asset, respectively defined by the evolution
\begin{align}\label{eq:model01}
  dS_t &= S_t \left( \mu dt + \sigma dW_t\right), \\
  dB_t &= r B_t dt,\label{eq:model02}
\end{align}
with $\mu$ the growth rate and $\sigma$ the volatility of the risky asset, $r$ the risk-free rate, and $dW_t$ the forward looking increment of a Wiener process. A self-financing trading strategy, which invests the fraction $\alpha$ in the risky asset and the fraction $(1-\alpha)$ in the risk-free asset, will result in a portfolio value which evolves according to the equation
\begin{equation}
  \frac{dP_t}{P_t} = \left[r+\alpha(\mu-r)\right]dt + \alpha\sigma dW_t.
\end{equation}
The analysis can be simplified by introducing the discounted relative portfolio value
\begin{equation}\label{eq:disc}
  \pi_t=\frac{P_t \exp(-r t)}{P_0}.
\end{equation}
 We can consider $\pi_t$ to be an index for the portfolio with $\pi_0=1$ and with $\pi_t-1$ the total return at time $t$ as measured in time zero dollars. Alternatively, if $r$ is the rate of inflation, it can be considered the real return. This definition eliminates the constant risk-free drift term $r dt$ from the evolution equation for the discounted portfolio value
\begin{equation}\label{eq:hjb:dPi}
  \frac{d\pi_t}{\pi_t} = \alpha(\pi_t,t)\left[(\mu-r) dt + \sigma dW_t\right],
\end{equation}
where we have indicated that the strategy may depend on the value of the portfolio as well as time. We are concerned here with problems for which the objective is the maximization or minimization of a terminal `reward' function $g(\pi_T)$. For a Kelly growth optimal portfolio this would be the period growth rate $g(\pi_T)=\log\pi_T$. The value function is defined as the expectation of the terminal reward over paths starting from the intermediate point $\pi$ at time $t$, and evolving according to (\ref{eq:hjb:dPi}) with the optimal control, i.e.\
\begin{equation}\label{eq:sc01}
  J(\pi,t) = \max_{\alpha}\mathbb{E}\left[\left. g(\pi_T)\right|\pi_t=\pi\right].
\end{equation}
It is significant that we have excluded a consumption type term in the objective, (a running cost/reward), as it would complicate the analysis considerably, but using the methods of appendix~\ref{app:legendre} such terms can be handled at least in simple cases (by adding the equivalent terms to (\ref{eq:lt06})).
The value function $J$ and the control $\alpha$ satisfies the Hamilton-Jacobi-Bellman equation
\begin{equation}\label{eq:HJB2}
  0=\partial_t J
  + \max_{\alpha(\pi,t)} \left\{\alpha(\mu-r)\pi\partial_{\pi} J
   +\tfrac{1}{2}\alpha^2\sigma^2\pi^2\partial_{\pi}^2 J
   \right\},
\end{equation}
which must be solved subject to the final time boundary condition
\begin{equation}\label{eq:HJB2cond}
  J(\pi,T) = g(\pi).
\end{equation}
The equations (\ref{eq:HJB2})-(\ref{eq:HJB2cond}) can also be used to find the strategy minimizing the expectation of a terminal objective $g(\pi_T)$ by replacing $\max$ by $\min$. In either case the objective function $g(\cdot)$ must have a form that makes the problem well-defined. As it stands the HJB problem defined by (\ref{eq:HJB2}) and (\ref{eq:HJB2cond}) is rather complicated as it involves two unknown functions $J$ and $\alpha$ linked by a combined PDE/maximization equation. The trick which renders the problem solvable is that under certain conditions we can perform the optimization over the control before solving for $J$, (see e.g.\ \citet{Oksen03,Yong99}). The condition is essentially that $J$ must have a shape which makes the optimization problem in (\ref{eq:HJB2}) have a well-defined solution. This must be confirmed a posteriori for the actual solution, (but in many applications it is rather obvious). In essence $J$ must be concave for a maximization problem and convex for a minimization problem. According to Bellman's principle, the optimal control must also be optimal over each sub-interval. In the present context this means that $\alpha$ must optimize (\ref{eq:HJB2}) at each instant of time. This is effected by setting $\partial_{\alpha}(\cdots)=0$, which gives the formal solution
\begin{equation}\label{eq:hjb:value:control}
  \alpha = -\frac{\mu-r}{\sigma^2}
  \frac{\partial_{\pi}J}{\pi\partial_{\pi}^2 J}.
\end{equation}
Substituting this back into (\ref{eq:HJB2}), the control is eliminated resulting in the HJB equation for the value function alone
\begin{equation}\label{eq:hjb:value}
  \partial_t J = \frac{(\mu-r)^2}{2\sigma^2}
  \frac{(\partial_{\pi}J)^2}{\partial_{\pi}^2 J}.
\end{equation}
The problem is then reduced to solving (\ref{eq:hjb:value}) subject to the terminal time boundary condition (\ref{eq:HJB2cond}) together with any additional boundary conditions (for example conditions specified on the edges $\pi^L\le \pi_t \le \pi^U$, which could be asymptotic). The standard procedure is thus to start by solving for the value function, and then as the second step to calculate the optimal strategy from (\ref{eq:hjb:value:control}).
A hint that (\ref{eq:hjb:value}) hides a simpler underlying structure is provided by the fact that a Legendre transform turns it into a linear second order PDE as discussed in appendix~\ref{app:legendre}.

\section{Nonlinear PDE for the optimal strategy}
\label{sec:nonlinPDE}
It is natural to ask, if there is a PDE for the optimal strategy without any reference to the value function. Such an equation would not only be interesting in its own, but would potentially allow us to circumvent the value function, and could have considerable practical benefits. In this section we show how it is possible to eliminate the value function, and derive a non-linear partial differential equation which is obeyed by the optimal control itself. The derivation requires that the functions involved are sufficiently smooth in the interior of the domain, but in practice this is not necessarily a great limitation.

The starting point is the HJB for the value function (\ref{eq:hjb:value}) and the equation for the optimal control (\ref{eq:hjb:value:control}). If we take the HJB equation (\ref{eq:hjb:value}) together with the equations which follow by operating on it with $\partial_{\pi}$ and $\partial_{\pi}^2$ and the equation for the optimal control (\ref{eq:hjb:value:control}) together with the equations which follow by operating on both sides by $\partial_{t}$, $\partial_{\pi}$ and $\partial_{\pi}^2$, then we have a system of seven equations for the unknowns $\alpha$, $\partial_t\alpha$, $\partial_{\pi}\alpha$, $\partial_{\pi}^2\alpha$, $\partial_t J$, $\partial_{\pi} J$, $\partial_{\pi}^2 J$, $\partial_t \partial_{\pi} J$, $\partial_t \partial_{\pi}^2 J$, $\partial_{\pi}^3 J$, $\partial_{\pi}^4 J$. Our aim is to derive a single equation only involving the stochastic control and its derivatives $\alpha$, $\partial_t\alpha$, $\partial_{\pi}\alpha$, $\partial_{\pi}^2\alpha$. On the surface it would look like we would need eight equations in order to be able to eliminate the seven derivatives of the value function and still be left with a single equation for the control, but note that the stochastic control as given by (\ref{eq:hjb:value:control}) only depend on the value function through the ratio between the two derivatives $\partial_{\pi}J/\partial_{\pi}^2 J$, which means that only six equations are needed to eliminate all the $J$'s. If we start by using four equations to eliminate $\partial_t \partial_{\pi} J$, $\partial_t \partial_{\pi}^2 J$, $\partial_{\pi}^3 J$, $\partial_{\pi}^4 J$, the three remaining equations can be put in the form
\begin{align}
  \partial_{\pi}^2 J &= -\frac{\mu-r}{\sigma^2}\frac{\partial_{\pi} J}{\pi\alpha}, \\
  \partial_{\pi}^2 J &=-\frac{\partial_t J} {\partial_t \alpha} \left[\pi^{-2}\partial_{\pi}(\pi^2 \partial_{\pi}\alpha)\right],\\
  \partial_{\pi}^2 J &=\frac{(\mu-r)^2}{2\sigma^2}
  \frac{(\partial_{\pi} J)^2}{\partial_t J}.
\end{align}
This gives us two equations for the ratio $\partial_{\pi} J/ \partial_t J$, respectively
\begin{align}
  \frac{\partial_{\pi} J}{\partial_t J} &=-\frac{2}{(\mu-r)\alpha\pi},\label{eq:alpha2value} \\
  \frac{\partial_{\pi} J}{\partial_t J} &=\frac{\sigma^2\pi\alpha}{(\mu-r)\partial_t\alpha}\left[\pi^{-2}\partial_{\pi}(\pi^2 \partial_{\pi}\alpha)\right].
\end{align}
It follows that the optimal strategy must obey the equation
\begin{equation}\label{eq:alpha:pde}
  \partial_t\alpha = -\frac{\sigma^2}{2}\,\alpha^2\partial_{\pi}\left(\pi^2\partial_{\pi}\alpha\right).
\end{equation}
This is a non-linear PDE which is second order in the portfolio value and first order in time. The non-linearity is manifested in the $\alpha^2$ factor on the right hand side. Like the HJB for the value function, it must be solved backwards in time from a final time boundary condition. In general the solution must also obey boundary conditions in the $\pi$-direction which may be asymptotic. It is reminiscent of a non-linear diffusion equation in reverse time, but it does not strictly have the required structure (see appendix~\ref{app:legendre}). The equation is independent of the effective drift rate $\mu-r$, but dependency on this quantity may enter via the boundary conditions. Since any constant will solve (\ref{eq:alpha:pde}), it is of little use for problems with such a solution. Rather (\ref{eq:alpha:pde}) is applicable to problems requiring a non-trivial solution in the sense of an optimal strategy which is a non-constant function of the portfolio value and time. The second caveat is that we must be able to establish the boundary conditions for the control for example from limit or asymptotic properties which the solution must obey. For some problems it may not be obvious how to find these boundary conditions in which case it will be necessary to revert to the HJB for the value function. If, on the other hand, we have a strategy solving (\ref{eq:alpha:pde}), then the corresponding value function can be found from the first order equation (\ref{eq:alpha2value}) from which it is determined up to a constant.

The equation (\ref{eq:alpha:pde}) can be transformed into different forms which may be more or less convenient for a particular investment problem. Some examples are discussed in appendix~\ref{app:transf}. An explanation for the form of the equation, and the nonlinearity in particular, can be found as follows. If we describe the optimal strategy by the amount invested in the risky asset rather than by the fraction of the portfolio value, i.e.\ using $\gamma(\pi,t)=\pi\alpha(\pi,t)$, then the equation takes the form
\begin{equation}\label{eq:gamma:pde}
  \partial_t\gamma = -\frac{\sigma^2}{2}\gamma^2\partial_{\pi}^2\gamma.
\end{equation}
Equation (\ref{eq:gamma:pde}) can also be obtained directly from the Legendre transformed HJB problem as discussed in appendix~\ref{app:legendre}.
Let $s_t$ denote the discounted risky asset price and $b_t$ the discounted bond price as per (\ref{eq:model01})-(\ref{eq:model02}) and (\ref{eq:disc}). Then $ds_t/s_t=(\mu-r)dt+\sigma dW_t$ and $db_t=0$, and the discounted portfolio value evolves according to the prescription
\begin{equation}\label{eq:port:gamma}
  d\pi_t = \gamma(\pi_t,t)\frac{ds_t}{s_t}.
\end{equation}
The process (\ref{eq:port:gamma}) is self-financing and is a special case of the general self-financing feedback strategy
\begin{equation}\label{eq:selffinance}
  d\pi_t = \phi(\pi_t,s_t,b_t,t)ds_t + \psi(\pi_t,s_t,b_t,t)db_t,
\end{equation}
with the self-financing condition $s_t d\phi + b_t d\psi=0$, with $\pi_t=\phi s_t + \psi b_t$, where $\phi$ is the number of risky asset shares and $\psi$ the number of risk-free bonds in the portfolio at time $t$. Since $db_t=0$, $\psi$ does not contribute to the dynamics, but is a slave keeping track of the accounting of the cash generated/needed for the re-balancing $d\psi=-s_t d\phi/b_t$. In our case the portfolio (\ref{eq:port:gamma}) can be decomposed as $\pi_t = \gamma(\pi_t,t) + \delta(\pi_t,t)$, with $\gamma=\alpha\pi_t=\phi s_t$ the value of the risky assets, and $\delta=\psi b_t$ the value of the risk-free investment. Whereas the portfolio $\pi_t$ is self-financing by construction, the risky asset investment, $\gamma(\pi_t,t)$, does not alone represent a self-financing quantity because a general self-financing strategy can redistribute the portfolio freely between $\gamma$ and $\delta$ as long as it keeps the sum unchanged. Because we are focussing on smooth strategies here, we can link the increment $d\gamma$ to the increments in the portfolio value, $d\pi_t$, and time, $dt$, via the expansion
\begin{align}\label{eq:gamma:expand}
  d\gamma &= (\partial_t\gamma)dt+(\partial_{\pi}\gamma)d\pi_t
  +\frac{1}{2}(\partial_{\pi}^2\gamma)(d\pi_t)^2 \nonumber\\
  &=\left[\partial_t\gamma
  +\frac{\sigma^2}{2}\gamma^2 \partial_{\pi}^2\gamma\right]dt
  +(\partial_{\pi}\gamma)d\pi_t.
\end{align}
It follows that when $\gamma$ represents an optimal trading strategy obeying (\ref{eq:gamma:pde}), this reduces to the expression
\begin{equation}\label{eq:gamma:optimal:redux}
  d\gamma = (\partial_{\pi}\gamma)d\pi_t =\gamma(\partial_{\pi}\gamma)\frac{ds_t}{s_t}.
\end{equation}
This means that although $\gamma$ is not self-financing for a general sub-optimal trading strategy, it will, for the optimal trading strategy, behave as a self-financing quantity with the total increment in $\gamma$ (from market evolution plus re-balancing) directly proportional to the increment in the portfolio (which because the portfolio is self-financing is given by market evolution alone). The same holds for the risk-free part of the portfolio which for the optimal strategy obeys $d\delta=(1-\partial_{\pi}\gamma)d\pi_t$. We note in particular that the nonlinearity of (\ref{eq:alpha:pde}) and (\ref{eq:gamma:pde}) is a consequence of the feedback nature of the portfolio evolution as specified by equation (\ref{eq:port:gamma}), and that the optimal strategy displays a certain balance between the risky and risk-free parts of the portfolio by only redistributing between these in proportion to the total increment in the portfolio.

\section{Known HJB solutions}
\label{sec:known}
In this section we comment on the application of the strategy equation (\ref{eq:alpha:pde}) to some standard investment problems, where there is a known analytic solution to the HJB equation. The purpose is to investigate to what degree (\ref{eq:alpha:pde}) can be applied with particular focus on the boundary conditions needed. Also \ref{sec:conststrat} and \ref{sec:kwts} establishes solutions which the optimal strategy in section~\ref{koswslr} must approach asymptotically. The examples have been chosen to illustrate problems where the optimal strategy is respectively constant, \ref{sec:conststrat}, value dependent, \ref{sec:kwts}, and dependent on both value and time, \ref{sec:brownestrat}.

\subsection{Kelly optimization and power function utility}
\label{sec:conststrat}

The free Kelly strategy is the unconstrained strategy maximizing the period growth rate and can also be interpreted as the strategy maximizing the terminal logarithmic  utility, i.e.\
\begin{equation}\label{eq:kellyUtil}
  J(\pi,t)=\max_{\alpha}\mathbb{E}\left[\left.\log(\pi_T)\right|\pi_t=\pi\right].
\end{equation}
The well-known optimal strategy, $\alpha=\alpha_K$, is to invest a constant fraction,
\begin{equation}\label{eq:kelly:alpha}
  \alpha_K= \frac{\mu-r}{\sigma^2},
\end{equation}
of the portfolio in the risky asset.
The corresponding value function is given by
\begin{equation}\label{eq:kelly:value}
  J(\pi,t)=\log\pi + \frac{(\mu-r)^2}{2\sigma^2}(T-t).
\end{equation}
Being constant (\ref{eq:kelly:alpha}) is a trivial solution to (\ref{eq:alpha:pde}).
A slight generalization of (\ref{eq:kellyUtil}) is the power function utility  $g(\pi_T)=(\pi_T)^{\eta}/\eta$, with $\eta<1$, which is also referred to by the acronym CRRA for constant relative risk aversion. The value function for this utility is given by
\begin{equation}\label{eq:mpp02}
  J(\pi,t) = \frac{\pi^{\eta}}{\eta}
  \exp\left(\frac{(\mu-r)^2\eta(T-t)}{2\sigma^2(1-\eta)}\right),
\end{equation}
and the corresponding optimal strategy is the constant
\begin{equation}\label{eq:mpp04}
  \alpha = \frac{\alpha_K}{1-\eta}.
\end{equation}
The free Kelly strategy corresponds to the special case $\eta\rightarrow 0$, and a fractional Kelly strategy, $\alpha=\kappa\alpha_K$, investing a constant fraction $\kappa$ of the free Kelly strategy, corresponds to optimizing a power function utility with $\eta=1-1/\kappa$. A rational investor would never invest more than the free Kelly strategy unless forced to do so by some constraint. This means that $\kappa$ would normally be less than one and $\eta$ therefore negative corresponding to a utility which is capped to the upside. Benchmarked fractional Kelly strategies have recently been investigated in detail by \citet{Davis08,Davis10}.

\subsection{Kelly with terminal stop}
\label{sec:kwts}

If an investor has a critical level $\pi_c<\pi_0$ below which his wealth must never fall, and therefore can only put at risk the part of his capital exceeding this level, then the capital below the critical level is in effect irrelevant to his utility. A Kelly investor will therefore optimize the growth rate of the capital exceeding the critical level. This goal is expressed in the value function
\begin{equation}\label{eq:kf01}
  J(\pi,t)=\max_{\alpha}\mathbb{E}\left[\left.\log(\pi_T - \pi_c)\right|\pi_t=\pi\right],
\end{equation}
with the HJB solution given by
\begin{equation}\label{eq:kf02}
  J(\pi,t) = \log(\pi-\pi_c)+\frac{(\mu-r)^2}{2\sigma^2}(T-t).
\end{equation}
The corresponding optimal strategy is
\begin{equation}\label{eq:kf03}
  \alpha(\pi)=\alpha_K\left(1 - \frac{\pi_c}{\pi}\right).
\end{equation}
This is the unconditional free Kelly strategy multiplied by the fraction that the excess capital constitutes out of the total capital. We observe that (\ref{eq:kf03}) is a solution to (\ref{eq:alpha:pde}) because it is independent of time and obeys $\pi^2\partial_{\pi}\alpha =\text{constant}$.
It is therefore completely defined as the time independent solution to (\ref{eq:alpha:pde}) subject to the boundary conditions $\alpha(\pi_C)=0$ and $\alpha\rightarrow\alpha_K$ for $\pi\rightarrow\infty$. We note that (\ref{eq:kf03}) is a `Constant Proportion Portfolio Insurance' (CPPI) strategy with the multiplier equal to the free Kelly strategy, i.e.\ the amount invested in the risky asset is $\gamma(\pi)=\alpha_K(\pi-\pi_C)$.
More generally we conclude that any time independent optimal strategy must take the form
\begin{equation}\label{eq:kf04}
  \alpha(\pi) = A + \frac{B}{\pi},
\end{equation}
for some constants $A$ and $B$.

\subsubsection{Kelly with maximum drawdown limit}
\label{GrossmanZhou}
\noindent\citet{Grossman93}, (see also \citet{Cvitanic95,Cvitanic97}), have solved the problem of finding the optimal strategy for an investor subject to a maximum drawdown rule for a general class of utility functions. The essence of their solution can in the present context be stated as follows. Investors who in accordance with the Kelly principle are seeking to maximize the long term real/discounted capital growth rate
\begin{equation}\label{eq:gz01}
  g = \lim_{T\rightarrow\infty}\frac{1}{T}\mathbb{E}\left[\log\pi_T\right],
\end{equation}
subject to the maximum drawdown constraint (with $0\le\lambda<1$ a given constant)
\begin{equation}\label{eq:gz02}
  \pi_t\ge\lambda m_t,
\end{equation}
where $m_t$ is the high-water mark
\begin{equation}\label{eq:gz03}
  m_t=\max_{s\le t}\pi_s,
\end{equation}
should follow the strategy of investing an amount equal to the free Kelly strategy times the excess over the moving stop level in the risky asset, i.e.\ $\gamma(\pi_t)=\alpha_K(\pi_t - \lambda m_t)$, or in other words, follow the strategy
\begin{equation}\label{eq:gz04}
  \alpha(\pi_t) = \alpha_K\left(1-\frac{\lambda m_t}{\pi_t}\right).
\end{equation}
We note the obvious similarity between (\ref{eq:kf03}) and (\ref{eq:gz04}), but refer to the references above for more information.

\subsection{Probability maximizing strategy}
\label{sec:brownestrat}

\citet{Browne99a,Browne99b, Browne00a} has shown that the problem of finding the strategy which maximizes the probability of reaching a specified real return target, $b>1$, in a finite time period, $0\le t \le T$, has the following solution. As argued by Browne, the probability that the target is hit at any time $t\le T$ is equivalent to the probability that the terminal wealth hits the target, because once the target is reached the probability maximizing strategy would be to invest exclusively in the risk-free asset. The value function for this problem is thus defined by
\begin{equation}\label{eq:browne:01}
  J(\pi,t) = \max_{\alpha}\text{P}\left(\left.\pi_T\ge b\right|\pi_t=\pi\right).
\end{equation}
With our notation Browne's solution to the problem is given by the value function (with $\pi\le b$)
\begin{equation}\label{eq:browne:02}
  J(\pi,t)=\Phi\left(\Phi^{-1}\left(\frac{\pi}{b}\right)
  +\frac{\mu-r}{\sigma}\sqrt{T-t}\right),
\end{equation}
and the corresponding optimal strategy is given by
\begin{equation}\label{eq:browne:03}
  \alpha(\pi,t)=\frac{1}{\sigma\sqrt{T-t}}\,\frac{b}{\pi}\,
  \phi\left(\Phi^{-1}\left(\frac{\pi}{b}\right)\right)
  =\frac{1}{\sigma\sqrt{T-t}}\,\frac{\phi(\nu)}{\Phi(\nu)},
\end{equation}
with $\nu=\Phi^{-1}(\pi/b)$, and with $\phi$ the standard normal density and $\Phi$ the corresponding cumulated probability function. Straight forward differentiation will show that (\ref{eq:browne:03}) is indeed a solution to (\ref{eq:alpha:pde}). As shown in appendix~\ref{app:transf}, (\ref{eq:browne:03}) is an example of a solution which can be found from (\ref{eq:alpha:pde}) via separation of the variables. It would nevertheless have been difficult to arrive at (\ref{eq:browne:03}) directly from (\ref{eq:alpha:pde}), because although the boundary condition $\alpha(b,t)=0$ is obvious, this is not the case for other boundary conditions. The optimal strategy (\ref{eq:browne:03}) diverges in the limit $t\rightarrow T$ to such a degree that the probability of hitting the target level approaches a finite value as the remaining time goes to zero, i.e.\ $J(\pi,t)\rightarrow\pi/b$ for $0<\pi\le b$. A strategy with finitely capped leverage would have $J(\pi,T)=0$ for $\pi <b$.

\section{Kelly optimal strategy with stop-loss rule}
\label{koswslr}
\begin{figure}
\begin{center}
\begin{subfigure}[b]{0.35\textwidth}
  \includegraphics[width=\textwidth]{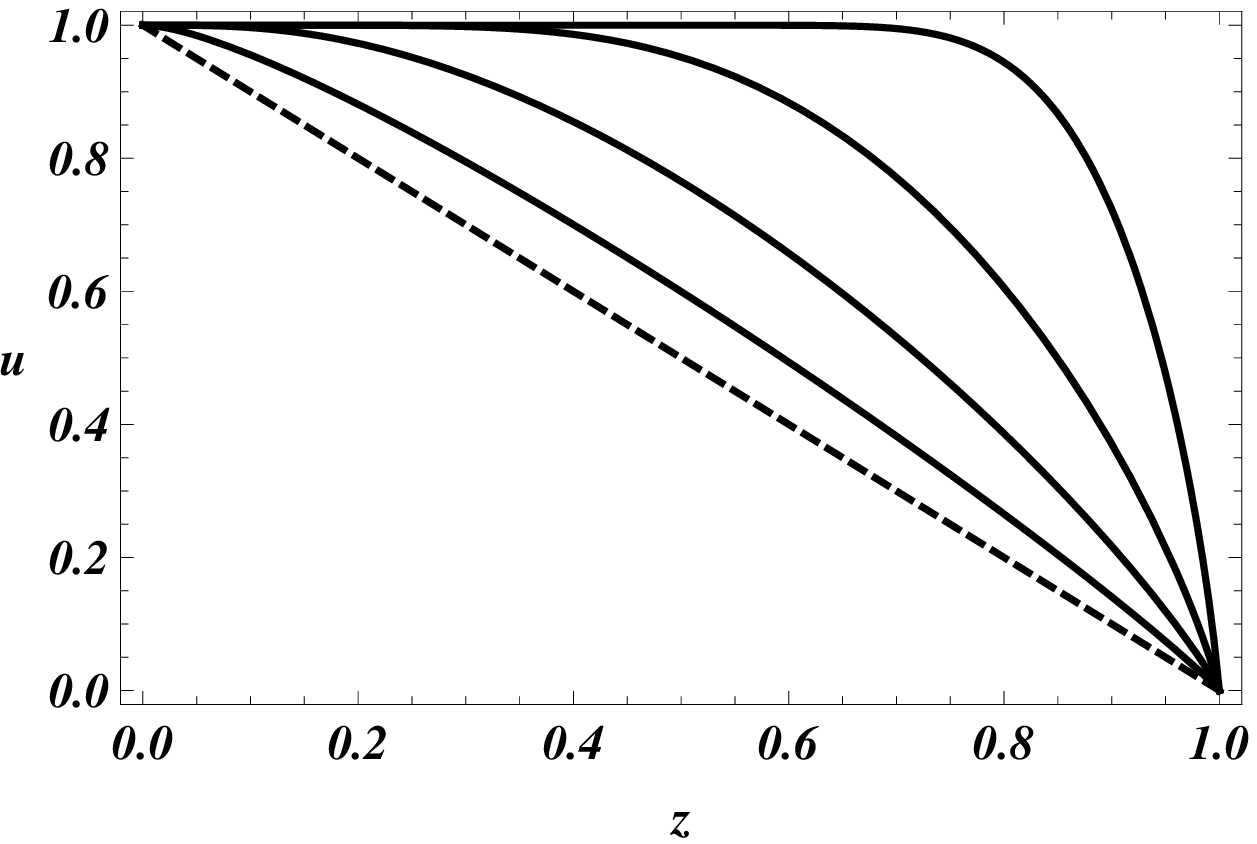}
  \caption{}
\end{subfigure}
\begin{subfigure}[b]{0.35\textwidth}
  \includegraphics[width=\textwidth]{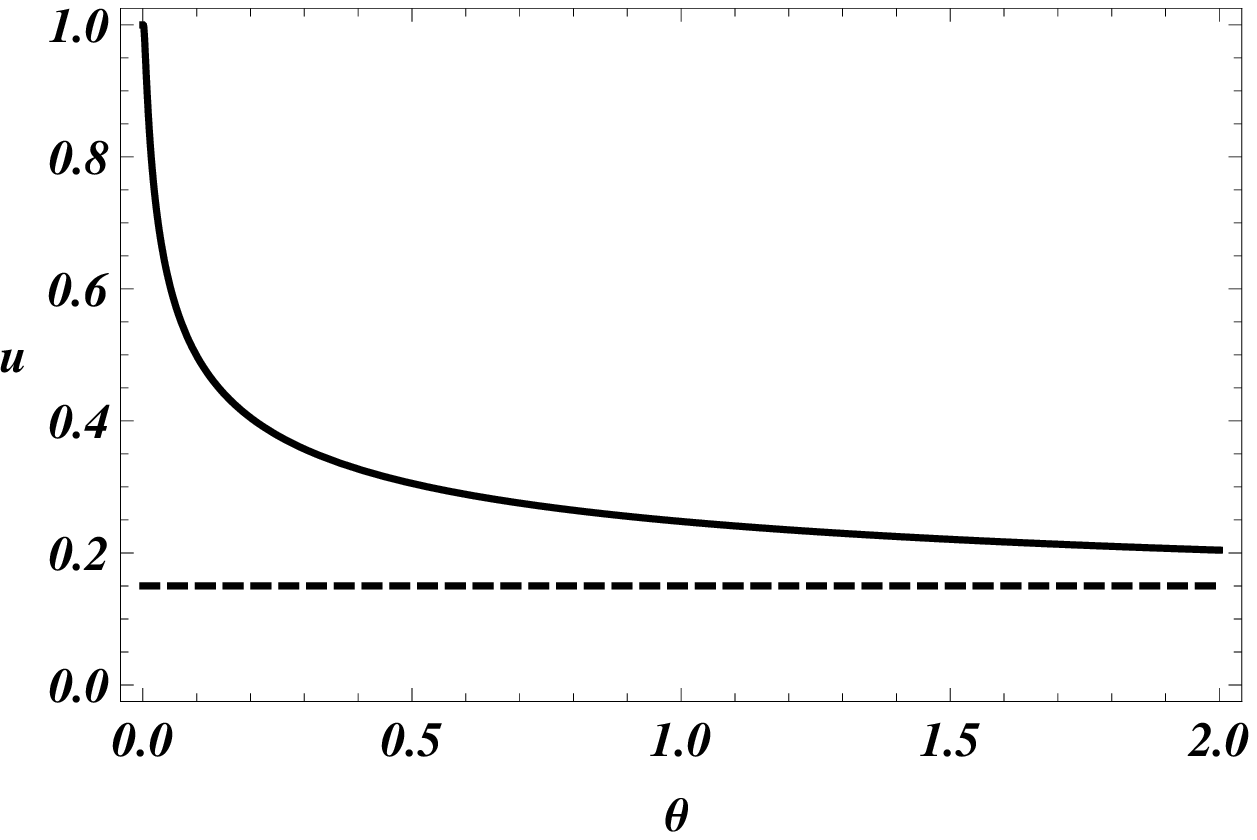}
  \caption{}
\end{subfigure}
\end{center}
\caption{\small (a) The scaled strategy $u(z,\theta)$ as a function of $z$ for four fixed values $\theta=0.01$, $0.1$, $0.5$ and $2.0$.
(b) The scaled strategy $u(z,\theta)$ as a function of $\theta$ for $z=0.85$ corresponding to a stop-level $15\%$ below the portfolio value. In both graphs the dashed line shows the asymptotic limit for $\theta\rightarrow\infty$.}\label{fig:asymp}
\end{figure}

We consider the investment problem of finding the strategy which optimizes the long term growth rate when investment is subject to a periodically reset stop-loss rule. To study this problem we consider the minimal model introduced in (\ref{eq:model01})-(\ref{eq:hjb:dPi}) with an investment universe consisting of one risky and one risk-free asset. The stop-loss rule means that if in a given period the value of the portfolio $\pi_t$ drops to a certain level $\pi_c$ (the stop-level), then the risk limit drops to zero, and the portfolio must be invested exclusively in the risk-free asset for the remainder of that period, and therefore the portfolio value will remain at $\pi_c$. At the beginning of a new period of duration $T$ the stop-loss rule is reset, and the level $\pi_c$ is set a fixed number of percentage points below the opening portfolio value $\pi_0$. In this model consecutive investment periods are therefore both independent and identical in distribution. We seek a strategy defined on the domain $(\pi,t)\in [\pi_c,\infty]\times [0,T]$. Since we know that the optimal strategy must solve the HJB problem and in particular the strategy PDE (\ref{eq:alpha:pde}), it follows that it must be smooth in the interior of the domain of definition, and that it is completely determined by the boundary conditions. It is intuitively clear that in the limit where the portfolio value is far above the stop-level, the stop is irrelevant and the optimal strategy will equal the free Kelly strategy, i.e.\ $\alpha(\pi,t)\rightarrow\alpha_K$ for $\pi/\pi_c\rightarrow\infty$. On the other hand, the stop-rule means that the optimal strategy must equal zero along the lower boundary on the portfolio value, i.e.\ $\alpha(\pi_c,t)=0$ for all $t\in [0,T]$. Finally it is clear that sufficiently near the final time boundary as $t\rightarrow T$ with a portfolio value strictly above the stop-level, $\pi>\pi_c$, the optimal strategy will approach the free Kelly strategy with the limit $\alpha(\pi,T)=\alpha_K$ for $\pi>\pi_c$. This follows from the fact that the underlying process is continuous in time, and therefore in the limit $t\rightarrow T$ the stop-rule becomes irrelevant for a portfolio with finite leverage when the portfolio value is strictly above the stop-level, because there is not time enough left for the portfolio value to drop to the stop-level. To summarize we have the three Dirichlet boundary conditions
\begin{align}
  \alpha(\pi,t)\rightarrow\alpha_K, & \;\;\text{ for }\pi\rightarrow\infty,
  \text{ for any }t\in [0,T], \\
  \alpha(\pi_c,t)=0, & \;\;\text{ for }t\in [0,T],\\\
  \alpha(\pi,T)=\alpha_K, & \;\;\text{ for }\pi>\pi_c.
\end{align}
To solve the problem we employ the strategy laid out in (\ref{eq:transf01})-(\ref{eq:transf04A}) in appendix~\ref{app:transf}, i.e.\ change variables from $(\pi,t)$ to $(z,\theta)$ and use the scaled optimal strategy $u=\alpha/\alpha_K$. The new variables are respectively the scaled reciprocal portfolio value $z=\pi_c/\pi$ and the scaled time to period end $\theta=(T-t)/\tau$ with $\tau=2\sigma^2/(\mu-r)^2$. The domain of definition for the variables is $(z,\theta)\in [0,1]\times [0,\infty]$, where $z=1$ is the stop-level, $z=0$ corresponds to the limit $\pi/\pi_c\rightarrow\infty$, and $\theta=0$ is the final time boundary. The equation to solve is thus (\ref{eq:transf04A}) subject to the boundary conditions
\begin{align}\label{eq:ubound01}
  u=1, & \;\;\text{ for }z=0, \\
  u=0, & \;\;\text{ for }z=1,\\
  u=1, & \;\;\text{ for }\theta=0.\label{eq:ubound03}
\end{align}
These boundary conditions determine the solution uniquely, but as discussed in appendix~\ref{app:transf}, the asymptotic behavior along the fourth boundary, which is the limit $\theta\rightarrow\infty$, also follows. In this limit $(T-t)/\tau\gg 1$ and the opportunity cost of hitting the stop-level is prohibitively high. The optimal strategy cannot have an explicit dependence on time in this limit, and must therefore approach (\ref{eq:kf03}) which in the transformed variables reads $u(z,\theta)\rightarrow 1-z$ for $\theta\rightarrow\infty$.
We are not aware of an analytic solution to (\ref{eq:transf04A}) obeying the particular boundary conditions (\ref{eq:ubound01})-(\ref{eq:ubound03}), but solving it numerically is straight forward. The solution is well behaved and although more advanced methods can be used, since $0\le z^2 u^2\le 1$ the explicit Euler method, which can be implemented in just a few lines of code, is convergent for $\Delta\theta/(\Delta z)^2 \lesssim 0.5$.
\begin{figure}
\begin{center}
\begin{subfigure}[b]{0.35\textwidth}
  \includegraphics[width=\textwidth]{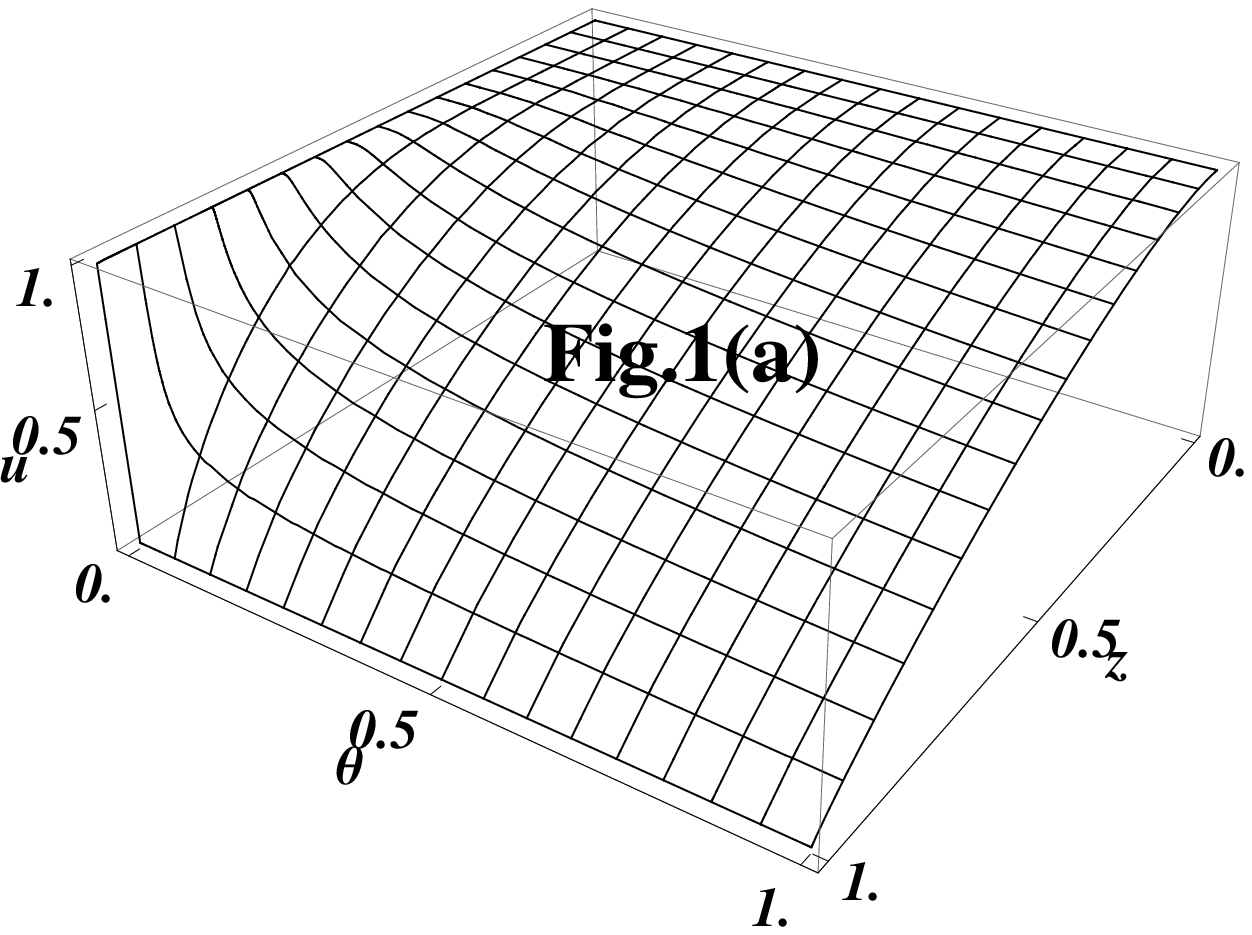}
  \caption{}
\end{subfigure}
\begin{subfigure}[b]{0.35\textwidth}
  \includegraphics[width=\textwidth]{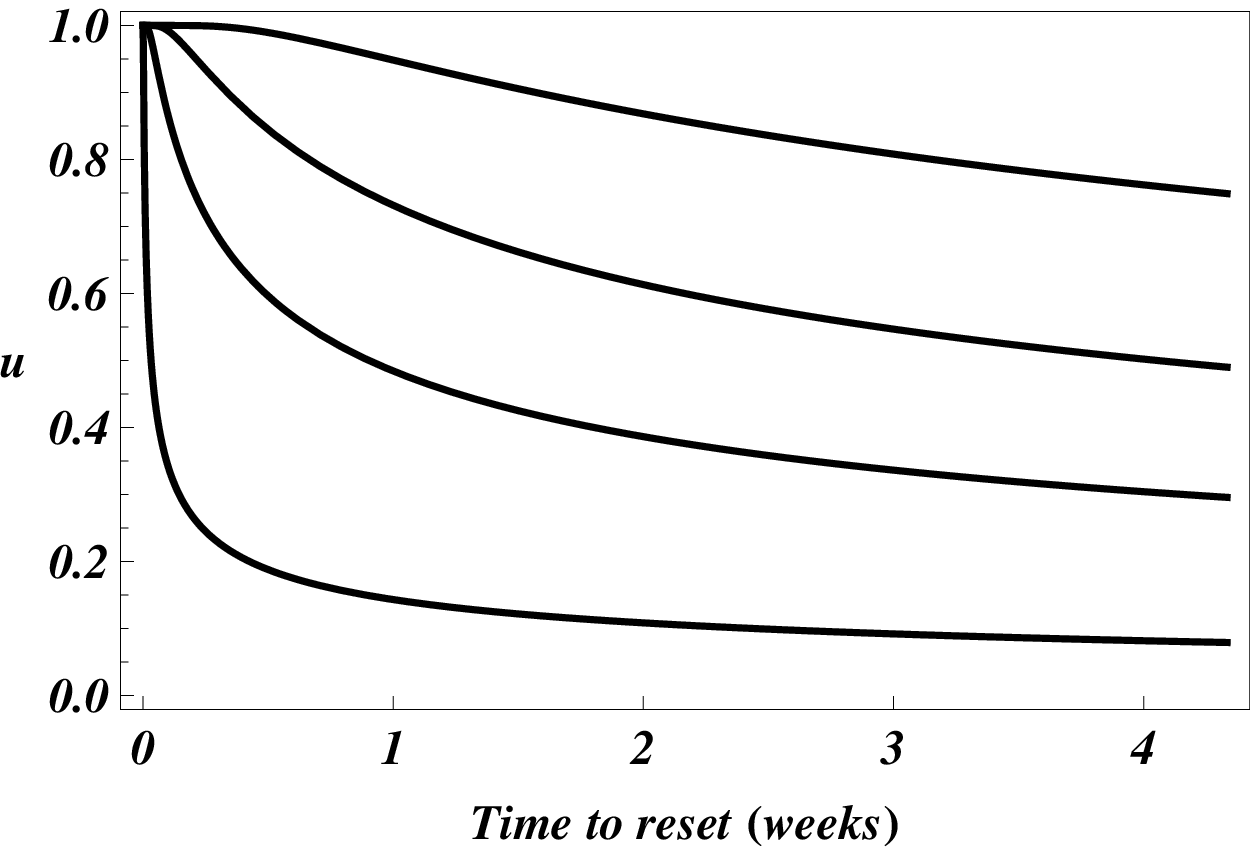}
  \caption{}
\end{subfigure}
\end{center}
\caption{\small (a) The generic scaled strategy $u(z,\theta)$ for $(z,\theta)\in [0,1]\times [0,1]$. The `dead zone' is indicated in red and the free Kelly limit in purple. (b) A specific example for an ex-ante Sharpe ratio of $s=1.0$ and a period of one month. The scaled strategy $u(\pi_{\delta},t)$ is plotted as function of time to period end for four different portfolio values specified by the distance to the stop level in percent. From the bottom the four curves correspond to stop levels which are respectively $\delta=1\%$, $5\%$, $10\%$, $20\%$ below the current portfolio value. Time to period end is indicated on the abscissa in weeks.}\label{fig:ustrat}
\end{figure}
Figure~\ref{fig:ustrat}(a,b) and \ref{fig:asymp}(a,b) show results calculated with this method. \ref{fig:ustrat}(a) shows the function $u(z,\theta)$, and \ref{fig:ustrat}(b) shows examples of $u(\pi_{\delta},t)$ for four fixed distances above the stop level. The ex-ante Sharpe ratio is $s=(\mu-r)/\sigma=1.0$, and the four curves correspond to stop levels which are respectively $\delta=1\%$, $5\%$, $10\%$ and $20\%$ below the current portfolio value, (i.e.\ $\pi_{\delta}=\pi_c/(1-\delta)$ and $z_{\delta}=1-\delta$). The time period in \ref{fig:ustrat}(b) is one month corresponding to a fund with monthly stop-level reset, and is indicated on the abscissa in weeks remaining until the reset.
Figure~\ref{fig:asymp}(a,b) illustrates the asymptotic convergence towards the terminal stop strategy as $\theta$ increases.

The main parameter setting the scale for the solution is the ex-ante Sharpe ratio of the risky asset $s=(\mu-r)/\sigma$. The optimal fraction $u=\alpha/\alpha_K=\alpha\sigma/s$ only depends on the distance to the stop-level through $z=\pi_c/\pi$ and the time left through $\theta=(T-t)/\tau$. The characteristic time scale $\tau=2/s^2$ is determined by the ex-ante Sharpe ratio. The free Kelly leverage is given by the ratio between the Sharpe ratio and the underlying volatility, i.e.\ $\alpha_K=s/\sigma$, but the volatility of the free Kelly portfolio equals the Sharpe ratio, i.e.\ $\alpha_K\sigma =s$. In practice other constraints are likely to kick in before the leverage reaches the free Kelly limit $u=1$, and the leverage will then be capped below this limit. If the portfolio is subject to a VaR limit corresponding to a maximum volatility $\sigma_{\text{max}}$, then $\alpha\sigma=u s\le\sigma_{\text{max}}$, and the scaled strategy is subject to the cap $u\le\sigma_{\text{max}}/s$.
With an ex-ante Sharpe ratio of $s=1.0$ as in figure~\ref{fig:ustrat}(b), a portfolio with a $3\%$ VaR limit at $95\%$, 1~day, corresponding roughly to $30\%$ annual volatility, would for example be subject to the cap $u\le 0.3$.

\subsection{Application to portfolio of multiple assets}

Consider a portfolio with one risk-free and multiple risky assets each evolving according to the standard model of geometric Brownian motion
\begin{equation}\label{eq:multi01}
  dS^k_t = S^k_t\left(\mu_k dt + \sigma_k dW^k_t\right).
\end{equation}
If $\alpha_k$ denote the allocation to the $k$'th risky asset, then the discounted portfolio value will evolve according to
\begin{equation}\label{eq:multi02}
  \frac{d\pi_t}{\pi_t} = \sum_k \alpha_k\left((\mu_k-r)dt
  +\sigma_k dW^k_t\right).
\end{equation}
The free Kelly portfolio is the portfolio optimizing the objective expectation of the growth rate, and it follows from
\begin{equation}\label{eq:multi03}
  \mathbb{E}\left[d\log\pi_t\right]
  =\left(\boldsymbol{\alpha}^{\dagger}
  (\boldsymbol{\mu}-\boldsymbol{r})
  -\tfrac{1}{2}\boldsymbol{\alpha}^{\dagger}\boldsymbol{C}
  \boldsymbol{\alpha}\right)dt,
\end{equation}
(with vectors and matrices indicated in bold and $\dagger$ indicating transposition) that this portfolio is given by the expression
\begin{equation}\label{eq:multi04}
  \boldsymbol{\alpha}_K = \boldsymbol{C}^{-1}\left(\boldsymbol{\mu}-\boldsymbol{r}\right),
\end{equation}
where $\boldsymbol{C}$ is the covariance matrix $C_{kl}=\sigma_k\sigma_l\rho_{kl}$.
We can now consider the restricted optimization problem consisting in finding the optimal strategy for an investor who has decided only to invest in portfolios proportional to the free Kelly portfolio. This is a very restricted problem compared to the full multi-dimensional portfolio optimization, but it is  not entirely irrational, since we know that this portfolio possesses many optimal properties, see e.g.\ \citet{Breiman61}, \citet{Hakansson71} and also \citet{Merton90}, \citet{Browne00a,Browne00b}, \citet{Davis10}. The free Kelly portfolio can itself be considered a (risky) asset, and we are therefore back to the situation discussed in the previous section. The dynamics of the Kelly portfolio is given by
\begin{equation}\label{eq:multi05}
  \frac{dK_t}{K_t}=\mu_K dt + \sigma_K dW_t,
\end{equation}
with $\mu_K=\sigma_K^2 = (\boldsymbol{\mu}-\boldsymbol{r})^{\dagger}\boldsymbol{C}^{-1}
(\boldsymbol{\mu}-\boldsymbol{r})$, which means that the unconstrained Kelly optimal allocation into the Kelly portfolio is $\mu_K/\sigma_K^2 = 1$, as it must be for consistency. It follows that we can find the optimal strategy for a trader subject to a stop-loss rule by the same procedure as for a single risky asset, and that the combined strategy can be expressed as
\begin{equation}\label{eq:multi06}
  \boldsymbol{\alpha}(\pi,t) = u(\pi,t)\boldsymbol{\alpha}_K,
\end{equation}
where the optimal strategy $u(\pi,t)$, which obeys equation (\ref{eq:alpha:pde}) with $\sigma=\sigma_K$, is a factor between zero and one.

\section{Conclusion}
\label{sec:conclusion}
We have derived a non-linear partial differential equation for the optimal strategy which does not involve the value function. The derivation requires that both the value function and the strategy should be sufficiently smooth in the interior of the domain. The non-linearity of the equation for the optimal strategy is a result of the feedback nature of the portfolio evolution, and the equation expresses a certain balance between the increments in the risky and risk-free parts of the portfolio. When boundary conditions can be established independently, the equation can be used to find the optimal strategy, and we have applied this method to find the Kelly optimal strategy subject to a periodically reset stop-loss rule.

Successful discretionary portfolio management is as much art as science, and intangible qualities like `superior and unbiased judgement', `alertness to position and funding stresses', `mental resolution and confidence to act before the crowd', `superior information gathering and networking', etc., are at least as important as any quantitative method. Nevertheless, this does not mean that the quantitative side of optimal portfolio management is unimportant. The results derived in this paper can only provide a rough guide to real world portfolio optimization, mainly because they are based on a simplistic and benign model. The main limitation is the model of market randomness based on Brownian motion. This has the consequence that all variables are continuous in time without gap risk, and allow the portfolio composition to be adjusted after each infinitesimal market move. It is well known that real markets have more in common with Lev\'y{}-processes with index less than two, and display jumps (gaps), serial correlation (trends) and volatility persistence (clustering). Transaction costs, limited liquidity and slippage are other problems. There is also the problem that even if the model were reasonable, we would still have to rely on statistical estimates of the true underlying parameters $\mu$ and $\sigma$. Thus the optimal strategy should incorporate the corresponding Bayesian uncertainty. Therefore, the methods discussed here must be adapted accordingly before they can be applied. Nevertheless, the results presented provide information about the optimal strategy in the limiting case of a very well behaved market.

\appendix

\section{Derivation via Legendre transform of the HJB}
\label{app:legendre}

The HJB equation (\ref{eq:hjb:value}) can be significantly simplified by a Legendre transformation using as the new independent variable in place of $\pi$, the slope of the tangent to the value function along the $\pi$-direction. Let the transformed value function $K(p,t)$ be defined by
\begin{equation}\label{eq:lt01}
  J(\pi,t) + K(p,t) = p\,\pi,
\end{equation}
with $p=\partial_{\pi}J$ and inversely $\pi=\partial_p K$. The Legendre transformation requires that the equation $p=\partial_{\pi}J(\pi,t)$ has a unique solution for $\pi$, i.e.\ that the inverse function $(\partial_{\pi}J)^{-1}(p)$ exist at least in an interval around the point $p$ for each point in time $t$. In the present context where $J$ is a value function, this is unlikely to be a problem, since $p$, being the rate of increase with respect to $\pi$, is a meaningful quantity. The Legendre transform is its own inverse and the properties of the transformation follow from the rules of differentiation. In general the transformation is time-dependent, i.e.\ the value of $p$ corresponding to a fixed value of $\pi$ is a function of time: $(\pi,t)\mapsto (p(t),t)$. Differentiating $K(p,t)$ with respect to $t$, while keeping $\pi$ fixed, therefore gives $(dK/dt)_{\pi}=\partial_t K + (\partial_p K)(dp/dt)_{\pi}$, but applying this procedure to (\ref{eq:lt01}) results in the simple relation
\begin{equation}\label{eq:lt02A}
  \partial_t J + \partial_t K=0,
\end{equation}
because $\pi=\partial_p K$. Another useful relation which follow by differentiation is
\begin{equation}\label{eq:lt02B}
  \partial_{\pi}^2 J=\left(\frac{d p}{d\pi}\right)_t =\left(\frac{d\pi}{d p}\right)_t^{-1}=(\partial_{p}^2 K)^{-1}.
\end{equation}
Applying the rules of the transformation, it follows that the HJB equation (\ref{eq:hjb:value}) is equivalent to the following linear second order PDE for $K(p,t)$
\begin{equation}\label{eq:lt03}
  \partial_t K = -\frac{1}{2}\sigma^2\alpha_K^2 p^2 \partial_p^2 K,
\end{equation}
which must be solved subject to the transformed terminal time boundary condition, e.g.\ for the Kelly utility $J(\pi,T)=g(\pi)=\log\pi$, we would have $K(p,T)=1+\log p$, because at time $T$ we have $p=\partial_{\pi}g=1/\pi$. Let us denote by $\phi(p,t)$ the transformed function corresponding to the optimal investment $\gamma(\pi,t)=\pi\alpha(\pi,t)$, via the definition
\begin{equation}\label{eq:lt04}
  \phi(p,t) = \gamma(\pi,t).
\end{equation}
It then follows from (\ref{eq:hjb:value:control}) that
\begin{equation}\label{eq:lt05}
  \phi = -\alpha_K p\, \partial_p^2 K,
\end{equation}
and from (\ref{eq:lt03}) and (\ref{eq:lt05}) we have
\begin{equation}\label{eq:lt045}
  \partial_t K = \frac{\sigma^2}{2}\,\alpha_K p\,\phi.
\end{equation}
The transformed HJB problem is significantly simpler than the original problem, and if we operate on (\ref{eq:lt05}) with $\partial_t$ (with $p$ fixed) and on (\ref{eq:lt045}) with $\partial_p^2$ (with $t$ fixed), then we get two equations from which we can eliminate $\partial_p^2 \partial_t K$. The result is the following linear equation for the optimal investment
\begin{equation}\label{eq:lt06}
  \partial_t\phi = -\frac{\sigma^2\alpha_K^2}{2}\, \partial_p\left(p^2\partial_p\phi\right).
\end{equation}
(\ref{eq:lt06}) is a diffusion equation with time running backwards, (i.e.\ it is a special case of the general non-linear one-dimensional diffusion equation with the structure: $\partial_t\phi = \partial_p\left[D(p,\phi)\partial_p\phi\right]$). To derive the corresponding equation for $\gamma(\pi,t)$ we have to remember that there is an implicit time-dependence in $p$ when $\pi$ is kept fixed and therefore $\partial_t\gamma = \partial_t\phi + (\partial_p\phi)(dp/dt)_{\pi}$. Otherwise the manipulations are straight forward and the result is (\ref{eq:gamma:pde}).

\section{Transformations and analytical solutions}
\label{app:transf}

The non-linear PDE for the optimal strategy (\ref{eq:alpha:pde}) can be transformed in many ways. Here we focus on a few with potential applications. Let $\alpha_K=(\mu-r)/\sigma^2$ denote the free Kelly strategy, and let $u(\pi,t)$ denote the scaled control
\begin{equation}\label{eq:transf01}
  u = \alpha/\alpha_K.
\end{equation}
If we introduce the scaled dimensionless time variable
\begin{equation}\label{eq:transf02}
  \theta = \frac{1}{2}\alpha_K^2 \sigma^2 (T-t)=(T-t)/\tau,
\end{equation}
then $\theta\ge 0$ measures the time remaining in units of the characteristic time $\tau = 2\sigma^2/(\mu-r)^2$, and the equation for the optimal strategy $u=u(\pi,\theta)$ takes the form
\begin{equation}\label{eq:transf03}
  \partial_{\theta}u =
  u^2\partial_{\pi}\left(\pi^2\partial_{\pi}u\right).
\end{equation}
Let $\pi_c$ denote a portfolio value of particular interest. Depending on the setting this could for example be the starting value, a stop-loss, or a target level. We can then define the scaled reciprocal portfolio value by
\begin{equation}\label{eq:transf04}
  z = \pi_c/\pi.
\end{equation}
The optimal strategy $u=u(z,\theta)$ then obeys the equation
\begin{equation}\label{eq:transf04A}
  \partial_{\theta}u =
  u^2 z^2 \partial_z^2 u.
\end{equation}
This form is convenient for investment problems with a stop level $\pi_c$, because $\pi_c\le\pi\le\infty$ translates to $z\in [0,1]$. Since $u^2 z^2\ge 0$, the sign of $\partial_{\theta}u$ is given by the curvature $\partial_z^2 u$. Thus as $\theta\rightarrow\infty$ the equation will seek to straighten out any curvature present in the initial state of the control at $\theta=0$ to approach a final state with $\partial_z^2 u=0$ consistent with the boundary conditions in the $z$-direction. For an investment problem with a stop level and $z\in [0,1]$, this means that $u(z,\theta)\rightarrow u(z)=u_0+(u_1-u_0)z$ for $\theta\rightarrow\infty$, with $u_0$, $u_1$ the respective boundary conditions.
For an investment problem with an upper target level $b$ the reciprocal portfolio value is conveniently defined as $z=b/\pi$, and the domain $0<\pi\le b$ then translates to $z\in [1,\infty [$. In this case the asymptotic behavior as $\theta\rightarrow\infty$ for a finite value of $z$ would be to approach a constant, $u(z,\theta)\rightarrow u_1$, which equals the boundary value of the control at the target level.
The equation (\ref{eq:transf04A}) for the optimal strategy has separable solutions of the form
\begin{equation}\label{eq:transf07}
  u(z,\theta) = \frac{f(z)}{\sqrt{\lambda\theta +c}},
\end{equation}
with constants $\lambda$ and $c$, and where the function $f(z)$ must solve the ODE
\begin{equation}\label{eq:transf08}
  f''(z)+\frac{\lambda}{2 z^2 f(z)}=0.
\end{equation}
Browne's probability maximizing strategy (\ref{eq:browne:03}) belongs to this class. The transformed solution resulting from the change of variables and scaling can be written as $u(z,\theta)=f(z)/\sqrt{2\theta}$, with $z=b/\pi$, where $b$ is the target level. The function $f(z)$, which solves (\ref{eq:transf08}) with $\lambda=2$, is given by
\begin{equation}
  f(z) = z \phi\left(\Phi^{-1}(z^{-1})\right),
\end{equation}
where $\phi$ and $\Phi$ refer to respectively the PDF and CDF of the standard normal distribution.

Alternatively, (\ref{eq:transf03}) can be simplified in analogy with (\ref{eq:gamma:pde}) by the transformation $w =\pi u=\gamma/\alpha_K$, which gives the PDE
\begin{equation}\label{eq:transf06}
  \partial_{\theta}w = w^2 \partial_{\pi}^2 w,
\end{equation}
for the scaled strategy variable $w=w(\pi,\theta)$. The equation (\ref{eq:transf06}) has self-similar solutions of the form $w(\pi,\theta)=F(x)$, where
\begin{equation}\label{eq:transf09}
  x=x(\pi,\theta)=\frac{\pi+a}{\sqrt{\lambda\theta+c}},
\end{equation}
for some constants $\lambda$, $a$ and $c$, and where the similarity function must solve the ODE
\begin{equation}\label{eq:transf10}
  F^2(x)F''(x)+\tfrac{1}{2}\lambda x F'(x)=0.
\end{equation}
\label{lastpage}
\end{document}